\RequirePackage{fix-cm}
\documentclass[twocolumn]{svjour3}          
\smartqed  
\usepackage{amsmath}

\usepackage{amsthm}
\usepackage{natbib}
\usepackage{graphicx}
\graphicspath{{./figures/}{../figures/}}
\usepackage{authblk}
\usepackage{subcaption}
\usepackage{booktabs}

\begin{document}
\renewcommand{\arraystretch}{1.3}

\title{A case for location based contact tracing}

\titlerunning{Location based contact tracing}        

\author{Atul Pokharel \and
Robert Soul\'{e} \and 
Avi Silberschatz
}

\authorrunning{Pokharel, Soul\'{e} and Silberschatz} 

\institute{A. Pokharel \at
              Wagner School of Public Service, New York University \\
              \email{pokharel@nyu.edu}           
           \and
           R. Soul\'{e} \at
           Department of Computer Science, Yale University
           \and
          A. Silberschatz \at
           Department of Computer Science, Yale University
}

\journalname{}
\date{November 20, 2020}

\maketitle

\begin{abstract}
  We present an evaluation of the effectiveness of manual contact tracing compared to bulletin board contact tracing.
  We show that bulletin board contact tracing gives comparable results in terms of the reproductive number, duration, prevalence and incidence but is less resource intensive, easier to implement and offers a wider range of privacy options.
Classical contact tracing focuses on contacting individuals whom an infectious person has been in proximity to.
A bulletin board approach focuses on identifying locations visited by an infectious person, and then contacting those who were at those locations.
We present results comparing their effects on the overall reproductive number as well as the incidence and prevalence of disease.
We evaluate them by building a new discrete time stochastic model based on the Susceptible Exposed Infectious and Recovered (SEIR) framework for disease spread.
We conduct simulation experiments to quantify the effectiveness of these two models of contact tracing by calibrating the model to be compatible with SARS-CoV-2.
Our experiments show that location-based bulletin board contact tracing can improve manual contact tracing.

\end{abstract}

\section{Introduction}

Contact tracing has been a large component of battling disease~\citep{cdccontact2020}. It works by identifying and alerting individuals who have been in contact with a congatious indivdual, so that they can isolate~\citep{hellewell2020feasibility}.
In the classical form of contact tracing, a contact tracer interviews the infected individual to help them remember who they may have contacted in the recent past.
The tracer then relays a message to isolate to each of these individuals.
This is the most common type of contact tracing and it is a manual process.
Recently, contact tracing has seen a large push from the tech community to see how they can help~\citep{apple2020}.
These more recent automated forms of contact tracing rely on technology to help the process of detecting contacts.
However, these proposals are essentially an automated form of classical contact tracing.
The main weakness of classical contact tracing is that it has a single point of failure: it depends entirely on the ability of an individual to accurately remember and identify people they contacted.
Analogously, the technology-assisted forms require a high degree of device adpotion.
Second, it requires the sharing and use of personally identifying information because the intention is to identify people uniquely.
Third, it assumes that enough of these contacts can be, or are willing to be, contacted out of the blue by the contact tracer that the spread can be slowed.

A different type of contact tracing is also used, although less frequently.
Countries like Israel have used a buleltin board style approach for contacting tracing.
Bulletin board contact tracing differs from the manual approach because it uses location-identifying information rather than individual-identifying information.
It consists of identifying locations that an infectious person has visited, and then disseminating that information anonymously using a public bulletin board rather than trying to uniquely identify and contact particular individuals.

This begs the question: how effective are the different forms of contact tracing?
There have been no evaluations of the effectiveness of bulletin board contact tracing compared to classical contact tracing.
Evaluating them using simulations is challenging without a natural way to conceptualize their components.
To address this challenge, we divide the contact tracing problem into two components: contact set reconstruction and message routing.
This enables us to implement simulations to compare manual contact tracing and bulletin board contact tracing.

We implement the simulation by developing a stochastic discrete-time model based on the Susceptible, Exposed, Infected and Recovered (SEIR) framework of disease spread.
We calibrate this model against what is known about SARS-CoV-2.
We then implement a prototype manual contact tracing method and a bulletin board contact tracing method.
And we compare their effectiveness in terms of the overall reproductive number, the incidence and the prevalence of disease. 

Our experiments show that bulletin board contact tracing performs similarly to manual contact tracing, or slightly wrose in some cases. It does not alter the overall reproductive number or the duration of disease compared to manual contact tracing. Bulletin board tracing reduces prevalence initially (Table~\ref{prevalence}), but then increases it. Incidence follows a similar trajectory.
Considered together with the efficiency gains from using a bulletin board approach to contact tracing, this suggests that bulletin board tracing can potentially improve the effectiveness of manual contact tracing.

Overall, the bulletin board approach is relatively easier to implement and requires many fewer resources than manual contact tracing. Even if the latter is automated it will likely still generate resistance for privacy reasons while also having accuracy issues.  That bulletin board contact tracing performs as well as (or slightly worse than) manual contact tracing, is a strong argument in its favor.  From a practical point of view, it is easier to implement and less resource intensive than manual tracing but gives comparable results, as we demonstrate with a new model.

\section{Types of Contact Tracing}

Contact tracing is a type of race against the spread of contagious disease~\citep{kaplan2003analyzing}. The disease spreads when one person infects at least one other. The number of people that a person directly infects under the assumption that everyone is equally susceptible is called the basic reproduction number($R_0$). Contact tracing is a method of finding out who has been exposed, and potentially infected by an infected individual. 
Once they are identified, they can be tested, treated and isolated to reduce the number of infections that would have otherwise occurred.
All forms of contact tracing proceed backwards in time from the discovery of this index individual to isolate other potentially exposed individuals. 
It works when this backward tracing is faster than the forward disease spread~\citep{kretzschmar2020impact,ivers2020can}.

Recently, there have been several public proposals for implementing contact tracing using technologies such as Bluetooth Low Energy (BLE). PACT~\citep{chan2020pact}, PEEP-PT~\citep{pepppt2020}, BlueTrace~\citep{bluetrace2020}, TCN~\citep{tcn2020}, DP-3T~\citep{dp3t2020} and Whisper~\citep{coalition2020} are among the most prominent. A major design goal of all of these proposals is user location-data privacy. Of these, PACT is the most developed in terms of implementation and testing. It uses Bluetooth enabled devices to determine when contact has occurred, and depends on how well the hardware permits assesing proximity.  Because many factors in addition to hardware features affect the accuracy of proximity detection, all implementations find it necessary to choose a mathematical model that will make this estimation with the desired level of accuracy.  Existing bluetooth hardware implementations cannot achieve greater than 60 percent accuracy in proximity detection~\citep{chan2020pact}, while GPS-based implementations have trouble differentiating different floors of the same building. Because of similar technological limitations and the privacy concerns with location tracking, none of these proposals  use GPS. Some contact tracing systems, such as the Whisper Protocol, allow users to report symptoms, as positive COVID-19 test results.
The PACT system proposes some solutions to potential issues with Bluetooth-based contact tracing having to do with infected surfaces. PACT proposes putting ``echoing'' devices on public surfaces that re-broadcast the chirps they collect from others. 

These implementations make certain assumptions about device adoption and use. The primary assumption is that the reporting and collection of location information is automatic, that is, the human does not input the locations. Instead, proximity and location is recorded by the device and considered to be more accurate than human assessments. All of these implementations assume that there is adequate adoption and that the device and the person are always close enough for the protocol to work. They also assume that people don't share devices, kids don't use their parents' phones even in the house etc. (i.e. the mapping from device to owner is not one to many). It is also not clear what happens when a person carries multiple id-generating devices, but they give conflicting assessments of proximity. Finally, the question of interoperability of these different protocols remains unaddressed. 
While we focus on location-based contact tracing, we do not make any assumptions about the specific implementation aside from separating the location recall and the message passing stages. The former can be achieved with the assistance of devices or not, and the latter requires a bulletin board system that is acessible to possible contacts. This generality permits implementations in which humans can correct previous reports that they posted about their locations.

These recent proposals are variations of classical contact tracing which take location into account in order to identify proximal individuals, find a route to them and pass a message to isolate.
These implementations remove humans from the loop of contact tracing i.e. they do not require that a contact tracer route a message to isolate to the contact as this is done automatically, nor do they require human intervention in deciding what information to share.
Ultimately, the bottleneck for these proposals is mobile device adoption.
Finally, the guarantees of trust are provided by device manufacturers for whom location tracking and information collection may be integral to their business model.

The bulletin board approach is entirely location-based, and need not be completely automated - this depends on the implementation.
For example, although they can use whatever location-based technologies they want to aid their memory, the infected individual (or the contact tracer) ultimately decides whether and what information to post.
No other information is collected, or shared about this individual for the purposes of contact tracing.
Similarly, it is not necessary to know who was in the vicinity of this infected individual: this is determined by the concerned individuals themselves.
The bulletin board approach requires only that a person who tests positive push an anonymous message to a public bulletin board that is then pulled by interested individuals.
As a result, it also offers more nuanced privacy options to tested individuals - they do not need to share any identifying information, nor is any identifying information necessarily collected.
And it allows for more nuanced human-in-the-loop implementations depending on context. As an added benefit, we suspect that the ability to calibrate privacy to ones preferences, in particular, is likely to improve cooperation with this approach compared to manual contact tracing. It is important to note, however, that our models suggest that bulletin board tracing may not perform as well as manual contact tracing towards the end of the disease lifetime.
\subsection{Classical (Manual) Contact Tracing}

Classical contact tracing begins with a contact tracer interviewing the infected individual. The interview focuses on mining the individual's memory of people they may have been in contact with in the recent past, a time range equal to the period of infectiousness of the disease. The interviews may cover a variety of topics, such the individual's recent travels, in the interest of jogging their memory of individuals they may have contacted. All of these individuals could have been potentially exposed. So, the tracer then attempts to contact them directly through available means to request taking measures to prevent further spread. Classical contact tracing assumes that this set of contacts can be identified accurately enough, and that they can be contacted fast enough, to win the race against forward spread. Aided by  communication technologies, this race seems more winnable than ever. 

Classical contact tracing has three major weaknesses. First, it has a single point of failure because it depends entirely on the ability and willingness of an individual to accurately identify who they contacted. 
Second, it requires the sharing and use of personally identifying information. Indeed, the more specific and individually identifiable the information the better for the purposes of manual contact tracing. Third, it assumes that enough of these contacts can be, or are willing to be, contacted out of the blue that the spread can be slowed.
These concerns have led to early reports of mixed effectiveness of classical contact tracing~\citep{Mueller2020} with half~\citep{jpost2020} of those contacted lying and two thirds~\citep{Mahase2020} refusing to cooperate.

\subsection{Bulletin Board Tracing}
\begin{figure}
    \centering
    \includegraphics[width=0.9\linewidth]{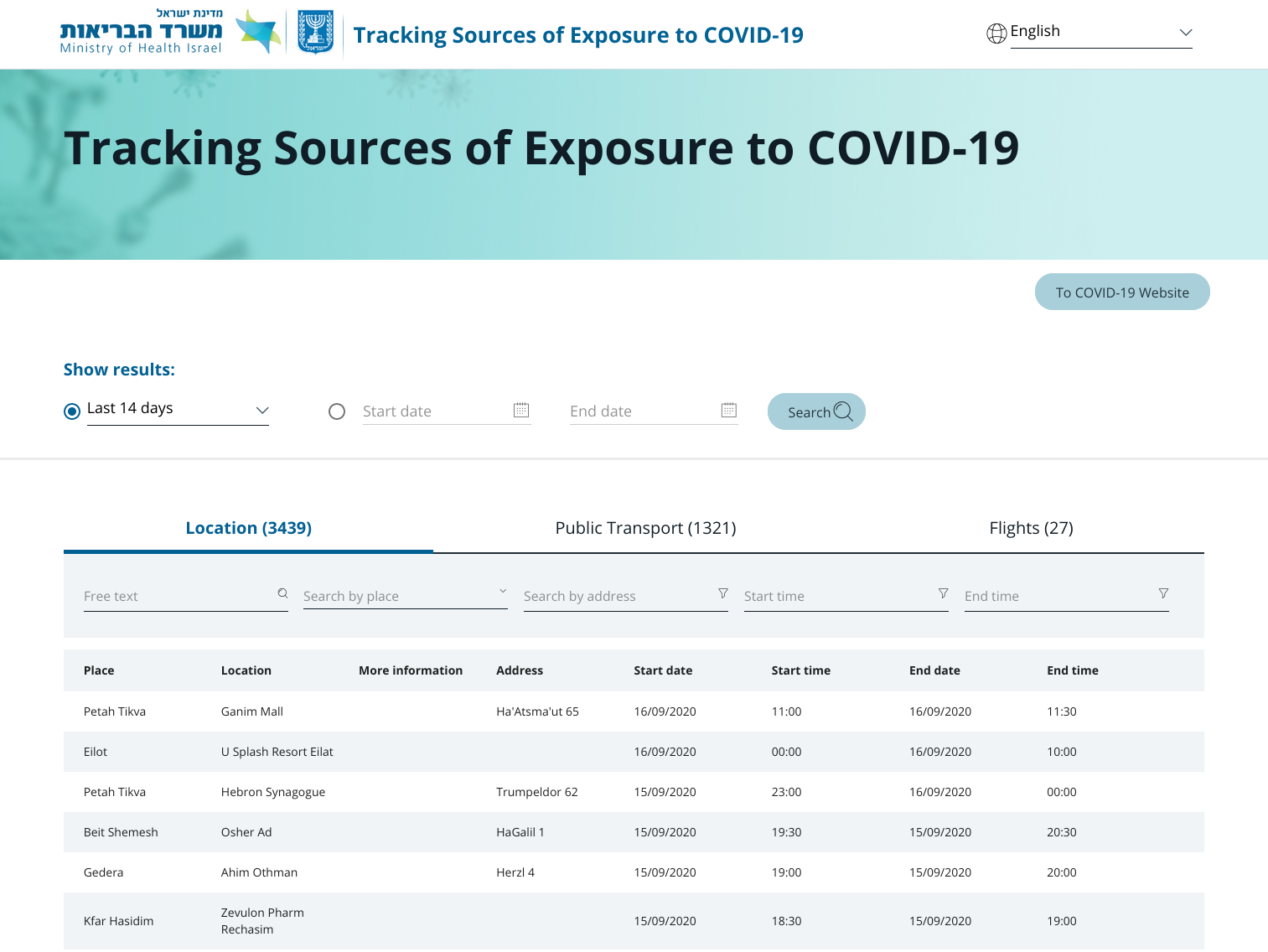}
    \caption{Israel's bulletion board for contact tracing. URL: https://coronaupdates.health.gov.il/corona-updates/grid/place  (Accessed 9/23/20 20:20)}
    \label{fig:israelbb}
\end{figure}

Bulletin board based contact tracing addresses some of the  problems with classical contact tracing. First, instead of reconstructing the set of individuals that a person has contacted, it relies on identifying the locations that they have visited. It does not rely on memories of individuals but on memories of locations. And second, instead of relying on the infected individual to identify people who may have been exposed, it relies on these people identifying themselves. More concretely, a message is publicly posted online that an infected person, without saying who, visited a particular set of locations. The general public are instructed to regularly check the bulletin boards for locations of interest to them. There are different ways to implement this, for instance with or without a contact tracing interview. And a variety of apps could be built on this platform to automate this process of checking or even posting messages. For example, Israel has a nationwide bulletin board system (Figure~\ref{fig:israelbb}).

Bulletin board contact tracing thus relies on the location memories of the infected individuals and those who may have been exposed.  Locations visited are typically easier to recall than people, many of whom are probably strangers. 
Next, because the message is broadcast, there is no need to route a message to a particular individual. This permits a wider range of privacy options both for the infected individual and their contacts.
Individuals choose which locations to receive messages about. So, personally identifiable information beyond what might be gleaned from these choices is irrelevant to bulletin board contact tracing. 
A major weakness of bulletin board tracing is that it relies crucially on the voluntary cooperation of a larger number of people than manual contact tracing. Beyond the cooperation of the the infected individuals, it additionally relies on the cooperation of all those who may have been infected. 
With the bulletin board approach, for instance, there is no notion of knocking on physical doors of private residences to convey a message: for it to work, these people will have to either frequently check the message boards or sign up to receive relevant alerts. 
Conceivably, an implementation of bulletin board contact tracing could automate both the posting of messages to the bulletin board and the pulling of messages from it. But this human-out-of-the-loop implementation would require accurate records of location history and a willingness to be tracked.

\section{Research Questions}

Our primary research question is, how does the effectiveness of manual contact tracing  at curtailing the spread of the SARS-CoV-2 virus compare to bulletin board contact tracing? For each of these two models of contact tracing, we further ask: (i) how efficient it has to be in order to be effective and (ii) what variables determine its effectiveness.

\section{Methodology}
We answer these questions using repeated agent-based simulations that combine a location-based model of disease spread with models of contact tracing. We use the CDC's ``best estimate'' scenario (Scenario 5) to calibrate this model~\citep{CDC2020}. As more is learned about real world spread, these parameters will need to be updated~\citep{gurdasani2020fallibility}. We define location as any physical space a person occupies for more than 15 minutes in a day. We conduct multiple simulations using a range of values for model parameters. We measure effectiveness in terms of (i) the time between first infection and the end of the last infection, (ii) the maximum number of people infected during this time period, and (iii) the average effective reproduction number.

\section{Mathematical Approach}

The problem of classical contact tracing can be separated into two problems: that of  reconstructing the set of contacts and that of routing the message that a person is in a contact set to them. Analogously, for bulletin board contact tracing these two problems are: the problem of reconstructing the set of locations and of relaying that a positive person had been to each location to others who were also there at that time.  These problems can be generalized as a reconstruction problem and a message routing problem. Our method to compare the performance characteristics of contact tracing under various assumptions using simulations relies on treating these two problems separately. From this perspective the two general types of contact tracing can be described as follows:

\begin{enumerate}
    \item \textbf{Manual:} The individual $i$ gets tested, if positive, a contact tracer asks them who they have been in contact with in the day range $R$, say $C_i(R)$, and contacts each of the individuals in the set.   The reconstruction problem consists of reconstructing the set of contacts for a relevant time period. This is done with the help of a contact tracer. The message passing problem consists of finding a way to contact these individuals and then  getting a message to them. Both of these are done by the contact tracer.
    \item \textbf{Bulletin board:} The individual $i$ gets tested, if positive anonymously posts their locations and test result to a bulletin board. All other individuals check this bulletin board to learn that they were in proximity. There are variations of this model with different privacy properties. Here the reconstruction problem consists of reconstructing the set of locations visited by the infected person over a relevant time period. This can be done with the help of a contact tracer, any physical or digital artifacts that one might have (such as receipts, photos or location histories) or independently. The message psssing problem consists of posting (pushing) an anonymous message to a bulletin board, again which can be done with assistance or independently. The other part of the problem is for the relevant individuals to ``pull'' relevant messages from the bulletin board. Note that it is not necessary for the contact tracer to find a route to the recipient.

\end{enumerate}

Two key parameters characterize the reconstruction problem and the message routing problem: recall accuracy of individuals and devices and the message routing method.

\subsection{Recall Accuracy (A)} 
Recall refers to the ability of a person to remember a past experience.
Recall accuracy ($A$) refers to the accuracy of a person's memory of a past experience. That is, how similar what they remember is to what was experienced. There are two types of recall relevant here: contact recall and location recall. The former refers to the ability to recall who one has come into contact with. The latter refers to the ability to recall where one has been.

In the classical model, the index agent's recall accuracy determines who gets a message. Index agent $i$'s contact recall is used to populate the set $C_i(R)$ and it determines accuracy of the set $C_i(R)$. 
Location recall of the index agent does not matter.
Recal accuracy of individuals in $C_i(R)$ may or may not be relevant. One way that it might be relevant is as a way to augment the agent's recall and make the set $C_i(R)$ more accurate. For example, for the contact tracer to ask the contacted person whether they remember being in contact with the index person.

In the bulletin board model, recall accuracy of the index agent $i$ as well as non-index agents determines who gets a message.
Here, the memory about index agent's location and non-index individuals location determines who gets a message. 
The contact recall of the index individual does not matter, but their location recall does.
Contact recall and location recall of non-index individuals also matters.
In human out of the loop implementations, recall accuracy of the machine determines who gets contacted since the recall of the machine is used to populate $C_i(R)$. A machine's recall accuracy, such as that of a mobile phone, is not guaranteed to be perfect.

Finally, the results of contact tracing can also be expressed in terms of the completeness of $C_i(R)$. For instance, the percentage in the set that are also in the real set is an indication of contact recall accuracy ($c$). Or, when we know that $C_i(R)$ is completely accurate (as in the case of flawless automation) then it is an indication of the effectiveness of message routing. We use the proportion of contacts recalled, and the proportion of recalled contacts who are reachable as parameters in our models. 

\subsection{Message routing method}
In the manual model of contact tracing, the message is passed to the correct recipient only through a contact tracer who must use available means to route the message. In the bulletin board model, the signal is effectively broadcast to all possible recipients by posting in a public location, and the recipients need to pull the messages and read those that are relevant to them.
If we assume that recall accuracy for all individuals is the same, then the two models of contact tracing are distinguished by the message passing mechanism. That is, they are distinguished by the way that the message reaches the elements of $C_i(R)$ from index agent $i$.

There are two processes within the signal routing mechanism that can alter its effectiveness:
\begin{enumerate}
    \item \textbf{route finding ($f$):} How one finds the route that the message should take to reach the destination, in this case an element of $C_i(R)$ for index agent $i$.
    \item \textbf{message passing ($p$):} The mechanism by which the message is passed to the destination, in this case an element of $C_i(R)$.
\end{enumerate}

For agent-based contact tracing, route finding is dominant because these individuals have to be tracked down after they are identified, but once they have been tracked down passing the message to them is easier. For bulletin board, message passing is more dominant because the message is pulled from the bulletin board by each individual. The contact tracer does not have to find a route and instead posts to one location. In other words, the tracer only has to find a route to the bulletin board which is trivial by design.
It is important to note that the processes for reconstructing the contact set and routing the message have to both be completed within a reasonable amount of time. In the case of SARS-CoV-2, an upper bound is 11 days because those to be contacted are infectious for that time.
So if the contact tracing processes take longer than that to implement, it is ineffective at stopping the spread of the virus.
A second constraint in manual contact tracing comes from the availability of contact tracers. We assume that there is always an adequate supply of contact tracers to be able to route the message to recalled individuals within the instantiation time. 

\section{Generalization}
 Let $T_c$ be the time between infection and the time that the message to get tested reaches an individual. 
Consistent with \cite{Chang2020.07.09.20149351} $T$ is the number of days before isolation and after infection. 
So, $T=T_c+T_E$ where $T_E>0$  because the steps following message receipt such as testing, getting the result and then entering isolation are not instantaneous. 

The models of tracing alter $T_c$. Hence,
\begin{equation*}
T_c=g(A, R, f, p)
\end{equation*}

For SARS-Cov-2, the relevant time period is 11 days ($R=11$). So,
\begin{equation} \label{tc}
T_c=g(A, f, p)
\end{equation}

\subsection{Early Transmission}
Under the assumptions of the standard SEIR model, a newly infected but not yet infectious person enters an exposed state for an exponentially distributed length of time $\tau^E$ with mean 1/$\mu_E$, after which they become infectious for an exponentially distributed duration $\tau^I$ of mean 1/$\mu_I$ during which transmission again occurs at constant rate $\beta$.

In the standard SEIR model describing an outbreak in a population of size N, the expected number of secondary infections generated by an infected person early in an outbreak, the reproductive number $R_0$, is given by

 \begin{equation} \label{rzero}
\begin{split}
R_0=N\beta/\mu_I
\end{split}
\end{equation}

where $\beta$ is the transmission rate per unit time. Given epidemiological estimates of the reproduction number, this enables the transmission rate to be written as~\citep{kaplan2003analyzing}.

\begin{equation} \label{b}
\begin{split}
\beta=R_0\mu_I/N
\end{split}
\end{equation}
 
In the agent based model to follow, each individual $i$ in the population is assigned the same initial transmission rate (and basic reproduction number $R_0$) as explained in Section~\ref{sec:CA} below. They end up with different effective reproduction numbers as a result of the simulation.

\subsection{Two Reconstruction Problems In Contact Tracing}
Before explaining the computational approach that we use, it is important to note that the reconstruction problem at the heart of of contact tracing can be formulated as centered on individuals or locations.

\subsubsection{Individual Based Reconstruction}
Let $\Omega$ be the set of all humans in a locality such that $|\Omega|=N$. Let $\omega_i$ be the subset ever contacted by human $i\in\Omega$. Given a time range R, $C_i(R)$ is the set of people contacted by $i$ in that range. Then, $C_i(R) \subset \omega_i$. Let $C^t_i(R)$ denote this set at time $t$.
Then, $\omega_i=\bigcup\limits_{t=0}^{t_L} C^t_i(R)$ where $t_L$ is the time of death. 
Next,  $\forall i,j$, $\omega_i \setminus \omega_j != \emptyset$, if one cannot be in contact with oneself. The recall problem for every person $i$ at time $t$ is the problem of assigning every other element $j \in \Omega$ a value in $\{0,1\}$ such that they are assigned a 0 if $j \notin C_i(R) $ and a 1 if it is. Next, let $C_i(R) \subset \Omega$  be the set of people $j$ such that $i \in C_j(R)$.  And, let $m_i(R) \subset \Omega$  be the set of people $j$ such that $i$ believes $i \in C_j(R)$ to be true. The reconstruction problem for each person $i$ is the problem of keeping $m_i(R)$ as similar to $C_i(R)$ as possible. To see why the reconstruction problem is non-trivial, consider the case where two people come into contact with each other but only one knows it (they may have their backs turned or eyes closed, or they might come in contact with a common surface in quick succession). At any given time $t$ and fixed $R$, three sets define each human being $i$: $C_i(R)$, $C^t_i(R)$ and $m_i(R)$. 

\subsubsection{Location Based Reconstruction}
Another way to formulate the reconstruction problem is to assume that every location records who has visited in every time period. That is, let $ml_{\{l,t\}}$ be the set of humans who have visited location $l$ in time period $t$. This might be the case, for instance, when a facial recognition device continuously records all who have visited a location.  
The reconstruction problem is the problem of reconstructing all $ml_{\{l,t\}}$.


\section{Computational Approach}\label{sec:CA}
As we have implemented it, there are three components  of the computational models of manual and bulletin board contact tracing. Two components are common to the models: the model of disease spread and the testing model. The third component, the model of contact tracing, is allowed to vary. This permits us to develop time-dependent agent based simulations to compute indicators of disease spread under the two contact tracing conditions without changing the models of disease spread and testing.
\subsection{The Disease Spread Model}
We use a location based SEIR-type model of spread that assumes homogeneous random mixing in the population and discrete time (in days). This type of mixing is appropriate for modeling local outbreaks, and represents a worst-case scenario.
Each community is assumed to consist of a population of size $N$ and a set of $L$ locations. We assume there is no importing of infections into the community.
The basic reproduction number $R_0$ over the population is held constant for each simulation, and every individual is thereby given a constant infection rate.
We assume that each individual's days in the exposed state ($\tau_i^E$) after infection is drawn from an exponential distribution with mean $1/\mu_E$.
We assume that each individuals days in the infectious state ($\tau_i^I$) after the exposed state is drawn from an exponential distribution with mean $1/\mu_I$.

Each individual is assigned an infection age, which represents the number of days since infection.
If the infection age is 0, they have never been infected. If (infection age)$\leq \tau_i^E $ then they are exposed but not infectious. If $ \tau_i^E < $ (age of infection)$\leq \tau_i^I$ then they are infectious. 
A person $i$ who is infectious can spread the infection to everyone in the same location who is not already infected. 
The number of people $n_i$ that $i$ actually infects in the simulation depends on $R_0$, $\tau_i^I$, and their movement between locations (and hence their total number of interactions).
Each time period, the total number of infections is counted.

\subsection{Population Mixing}
To enable a comparison of the buleltin-board approach, we model interactions using location assignment rather than stochastic pair formation and separation~\citep{kretzschmar1996measures}.
Each individual $i$ is assigned an integer value from $[1...L]$ using a discrete uniform probability distribution over the set.
All individuals assigned the same non-zero value are assumed to be in contact on day $t$.
The location assignment model has two desirable characteristics. First, the parameter determining disease spread can be specified in terms of location. And second, it incorporates the idea of movement in space. Both of these are important for modeling bulletin board contact tracing. As we show below, we calibrate the combined  population mixing and disease spread model to be equivalent to what is expected from more common mass-action models.

\subsection{Location Dependence}

In a location-based simulation, the exact mix of Susceptible ($S_t^l$), Exposed ($E_t^l$), Infected ($I_t^l$) and Recovered ($R_t^l$) people at each location is determined by the probabilistic model of population mixing.  The variables that characterize this model of mixing are the number of locations they can travel to ($L$), and the number of people ($N$). Hence, the effective reproduction number for an individual $i$ depends on these two variables.

\begin{equation} \label{locations}
    R_{0i}=f(L,N)
\end{equation}

\subsection{Approximating Mass-action Epidemic Models}
In common mass action models, $|S_{t}^l| \geq n_l^i$ and $L=1$ with homogenous random mixing and a constant rate of infection.
To generate an infection rate that is compatible with this in a simulation with $L>1$ and known $R_0$, we  assign each individual a constant scaled transmission probability.
To retain stochasticity, we draw the infectious and exposed durations from probability distributions.

Let the expected number of infections at time $t$ in our model be given by
\begin{equation}
\beta|S_{t}||I_{t}|=b\sum_{l=1}^{L}|S_{t}^l||I_{t}^l|
\end{equation}
Where
\begin{equation}
  |S_{t}^l|=|S_{t}|q_l
\end{equation}
and
\begin{equation}
  |I_{t}^l|=|I_{t}|q_l
\end{equation}

$q_l$ is the probability that the individual $i$ will be at location $l\in\{1,2..L\}$. $b$ is the scaled transmission rate and $\beta$ is the constant transmission rate in an ordinary SEIR model with a single location. Note that we are assuming that $b$ is also constant in our model, which is what permits us to pull it out of the sum. More generally, using (\ref{b}) we can rewrite the the scaled infection rate for an individual $i$ as
\begin{equation}
  b_i=\frac{\beta}{\sum_{l=1}^{L}q_l^2}
  =\frac{\mu_IR_{0i}}{N\sum_{l=1}^{L}{q_l^2}}
\end{equation}

In our experiments there is no heterogeneity in the underlying disease spread mdoel. So, $R_{0i}=R_0$ and does not vary at the individual level. In order to give everyone the same value of $b_i$, we use the mean value $1/\mu_I$ for $\tau_i^I$ above.

Since we assign an equal probability of being at any location, $q_l=1/L$, to each individual, we have, 
\begin{equation}\label{randommix}
  b_i =\frac{\mu_IR_{0i}L}{N}
  = \frac{\mu_IR_{0i}}{\rho}
\end{equation}    
 Where $\rho=N/L$ is the density of the model and $0\leq b_i \leq 1$.
 To spread the infection at location $l$, a susceptible person at location $l$ is infected with probability
\begin{equation}\label{eq:prob}
    p_l=1-\prod_{i\in I_t^l}(1-b_i)
\end{equation}
Thus, the expected number of infections at location $l$ is given by 
\begin{equation}
    n_t^l=|S_{t}^l|(1-\prod_{i\in I_t^l}(1-b_i))
\end{equation}
These new infections are then assigned at random to infectious people at the location of infection.
Note that (\ref{randommix}) makes explicit the effect of density on the transmission probability in our simulation.
Finally, it is now easier to see that if $L=1$, we have a mass action model with a single location.

Note that this formulation requires that $P\{|S_{t}^l| < n_l^t\}=0$.
To achieve this condition, it must be that $\mu_IR_{0i}\leq \rho$.
In our implementation, we choose simplicity over generality and fix $R_{0i}$ for each simulation, use $\mu_I$ in place of $\tau_i^I$ when computing the scaled transmission rate, and use a sampled $\tau_i^I$ to determine how many days an individual remains infectious. As we have calibrated the model below, $max(\mu_IR_{0i})<0.228$, and $min(\rho)=1$



\subsection{Testing Model}
Testing in this model assumes symptomatic testing because that is the type of testing that is associated with manual contact tracing. We assume that 60\% show symptoms. For those that show symptoms, we assume that the incubation period of the virus for each individual is drawn from a lognormal distribution with mean $\mu_P$ and standard devision $\sigma_P$, after whcih they start to show symptoms.
For simplicity, we assume that the individual's symptoms, if they are symptomatic, coincides with their infectious period.
We repeatedly sample the incubation period until we get a value in the individuals infectious period, $(\tau_E^i,\tau_i^I)$.
We assume that it takes every individual one day to recognize symptoms, and they do so accurately. There is a delay in receiving test results ($t_d$), and the test itself has a false negative rate of $P_N$ and a false positive rate of 0, as the false positive rate is only relevant to asymptomatic testing. We assume that individuals isolate immediately upon receiving a positive test result with probability $P_I$. 

\subsection{Implementation}
We implement the disease spread model using time steps of one day as follows:
\begin{enumerate}
\item Initialization: Begin with a population of susceptible people ($P$) 
and a number of locations($L$). Each individual($i$) is initialized with values for exposed duration, infectious duration, whether or not they are symptomatic and incubation period from the corresponding distributions (Table~\ref{tab:parameters}) to be applied in the event that they become infected. We also assign each person a transmission probability $b_i$ and infection age of 0 since everybody is uninfected.
\item Initial infection: At time $t_0$, we choose a randomly infected index person from the population. Initializing with more than one index person speeds up the infections, without changing the dynamics.
\item Movement: Every person is randomly assigned to one location with a uniform probability. In subsequent days, each location will have a mix of Susceptible, Exposed, Infectious and Recovered individuals. 
\item Generate infections: For each location, infect each susceptible person with the probability shown in Equation~\ref{eq:prob}. 
\item Repeat: Increase the infection age of each exposed or infectious individual by one day. Return to the movement step until the number of exposed and infectious people in the popultion is zero.
\end{enumerate}

The testing model is combined with this model of disease spread by adding the following steps to each time period before the Repeat stage.

\begin{enumerate}
\item Testing: Test each person in the population whose infection age is greater than their incubation period. Tests are ready after $t_d$ days.
\item Isolation: For each person whose results are ready and positive, isolate them with probability $P_I$. For each person who has been isolated for $R$ days, unisolate them. $R$ is the number of days after which an infected person is assumed to be no longer infected.

\end{enumerate}

Finally, the two different types of contact tracing are added on top of this by adding the following steps to be followed if a person $i$ tests positive.
For manual contact tracing:
\begin{enumerate}
\item Contact Reconstruction: Choose $P_c$ proportion of individuals from the set of people that $i$ has contacted in the last $R-S$ days. $S$ is the number of days an individual is assumed to be in the exposed but not infectious state.
\item Route Finding and Message Passing: Push a message to this set of contacts with some probability of reaching them ($P_r$).
\item Isolation: Isolate each person who receives the message with probability $P_{im}$.
  \end{enumerate}

For bulletin board contact tracing:
\begin{enumerate}
\item Location Reconstruction: Choose $P_c$ proportion of locations from the set of locations that $i$ has visited in the last $R-S$ days.
\item Message Passing: $P_s$ proportion of those in each of these locations receive a message that an infected person visited a relevant area. 
  \item Isolation: Each person who receives the message isolates with probability $P_{ib}$.
\end{enumerate}

\section{The four models}
 Four models are compared when $R_0$ is 2.5. Refer to Table~\ref{tab:parameters} for the typical values and symbols used.

\begin{itemize}
\item[\textbf{m0:}]\textbf{No Intervention:} This is the location assignment model of spread with no testing or contact tracing given a population size ($N$) and number of locations ($L$).
    The infectious period for each person is assumed to end after $\tau_I^i$ of infection, and assumed to start $\tau_E^i$ days after infection. $\tau_I^i$ is drawn from an exponential distribution with mean $1/\mu_I$ and $\tau_E^i$ is drawn from an exponential distribution with mean $1/\mu_E$. 
    The main parameters are:  \{$N$, $L$, $1/\mu_I$, $1/\mu_E$\}. The other models below build on this model.
    
    \item[\textbf{m1:}]\textbf{Testing Only:} Every individual is tested symptomatically. 40\% of the population is assumed to be asymptomatic. For others, symptoms arise after an incubation period drawn from a lognormal distribution with mean ($\mu_p$) 5.2 and standard deviation ($\sigma_p$) 3.9. $P_T$ proportion of people who show symptoms get tested. The result arrives $t_d$ days later with a false negative rate of $P_n$. People isolate with probability $P_I$ upon receiving a positive test result for $R$ days. The main parameters are: \{$N$, $L$, $1/\mu_I$, $1/\mu_E$, $t_d$, R, $P_T$, $P_n$, $\mu_p$, $\sigma_p$\}.
    
    \item[\textbf{m2:}]\textbf{Manual Contact Tracing:} Tracing begins  $T_m$ days after an individual receives a positive test. A contact tracer reconstructs $P_c$ proportion of the individual's contact set for the last $R-S$ days from the individual's memory, with or without assistance. The tracer then sends a message to all of these people and it reaches each target individual with probability $P_r$. Recipients isolate with probability $P_{im}$ upon receiving the message. It takes $T_m$ days for the manual tracing steps to be implemented. Typically these are securing cooperation, interviewing, assigning volunteers and locating recipients. We assume that $P_m$ proportion of infected individuals are willing to subject themselves to manual contact tracing.  The main parameters are 
    \{$N$, $L$, $1/\mu_I$, $1/\mu_E$, $t_d$, $P_T$, $P_n$, $\mu_p$, $\sigma_p$, $R$, $S$,  $P_c$, $P_r$,  $P_m$, $T_m$, $P_{im}$\}.
    
  \item[\textbf{m3:}]\textbf{Bulletin Board:} 
    An individual who receives a positive test posts a message to a bulletin board with probability $P_b$, which is the probability that they will cooperate with the requirements of bulletin board tracing. The message contains two pieces of information: (location, when visited). One message is sent to the bulletin board for each location they visited in their relevant location history. They do not share any personally identifying information. Cooperating individuals reconstruct $P_l$ proportion of their location history for the last $R-S$ days either alone or with assistance.  $P_s$ proportion of people overall are signed up to get bulletin board messages concerning any given location. The message to isolate reaches an identified person with probability $P_r$ in time $T_b$.  We assume that the recipients are willing to isolate with probability $P_{ib}$ when they receive a message this way.  The main parameters are  \{$N$, $L$, $1/\mu_I$, $1/\mu_E$, $t_d$, $P_T$, $P_n$, $\mu_p$, $\sigma_p$, $R$, $P_s$, $P_l$, $P_b$, $T_b$,$P_{ib}$\}.
\end{itemize}

\begin{table*}
\centering
\caption{Model parameters}
\label{tab:parameters}
\begin{tabular}{@{}p{1.5cm}p{1.5cm}p{2cm}p{11cm}@{}}
\hline
\textbf{Model} & \textbf{Parameter} & \textbf{Values} & \textbf{Description} \\
\hline
All   &     $N$      &   1000     &      Number of people.       \\
All   &     $L$      & 10       &    Number of locations. \\
All   &     $R_0$      & 2.5       &    Basic reproduction number \\
All   & $1/\mu_E$ & 3  & Mean of days exposed (Days immediately after infection before becoming infectious) for exponential distribution. \\
All   &     $1/\mu_I$ &      11    & Mean of days infected (Days after which an infected person is no longer infected) for exponential distribution.\\
m1,m2,m3   &     $\mu_p$ &      5.2    & Mean of the lognormal distribution of incubation periods.\\
m1,m2,m3   &     $\sigma_p$ &      3.9    &  Standard Deviation of the lognormal distribution of incubation period.\\
m1,m2,m3   &     $t_d$ &      1    & Testing delay: time between test and result.\\
m1,m2,m3   &     $P_I$  & 0.7          &    Probability of isolating after a positive test result.          \\ 
m1,m2,m3   &     $P_N$      &    0.2    &    Probability that the test will miss an infected person.  \\ 
m2, m3   &     $S$  &  3     & For tracing: days immediately after infection before becoming infectious. \\
m2,m3  &     $R$ &      14    & For tracing: Days after which an infected person is assumed to be no longer infected.\\
m2    &   $t_m$    &   1    &  Time taken to instantiate manual tracing for a person (such as by interviewing an infected individual)\\
m2    &   $P_c$    &    0.5   & Proportion of the contact set that a tracer is able to reconstruct.\\ 
m2    &   $P_r$    &    0.5   & Probability that a message reaches a person in the contact set.\\ 
m2    &    $P_m$    &   0.5    & Probability of positive person cooperating with manual contact tracing. \\ 
m2   &     $P_{im}$  & 0.7          &    Probability of isolating if asked by a manual tracer.          \\
m3    &   $t_b$    &    0   & Time taken to instantiate bulletin board tracing for a person.\\
m3    &   $P_l$    &    0.7   & Proportion of the location set that an individual remembers.\\ 
m3    &   $P_s$    &0.5       & Proportion of those who were exposed at a particular location who are subscribed to notifications or check the bulletin board daily.\\ 
m3    &    $P_b$    &   0.5    & Probability of positive person cooperating with bulletin board tracing.\\ 
m3   &     $P_{ib}$  & 0.7          &    Probability of isolating if notified by bulletin board tracing.          \\
\hline
\end{tabular}
\end{table*}

\section{Model Calibration}
\begin{figure}[htpb]
     \includegraphics[width=0.9\linewidth,height=6cm]{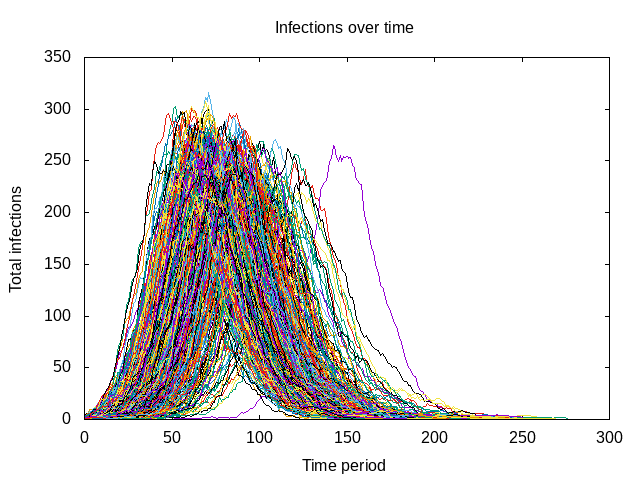}
     \caption{The sequence of infections of 1000 runs for calobrated base model (m0) with 1 location.}
     \label{fig:m0runs}
\end{figure}

The no intervention scenario is calibrated to be compatible with the currently known characteristics of SARS-CoV-2 as far as possible. When there is only a single location, the final sizes are compatible with what is expected from mass action SEIR models when $R_0$ is 1.5 (0.583), 2.25 (0.854) and 2.5 (0.89). Figure~\ref{fig:m0runs} shows the time traces of infection for each run of the simulation in this scenario.

\begin{table*}
\centering
\caption{\textbf{Calibration}: The base model (m0) with a single location and no heterogeneity. $R_0=2.5$, $N=1000$, $\rho=100$, 1000 runs.}
\label{m0calibrate}
\begin{tabular}{@{}rrrrrrr@{}}
\hline
  & Min.  & 1st Qu. & Median & Mean  & 3rd Qu. & Max.   \\
  \hline
  Duration of infection (days) & 11  & 148   & 176  & 149 & 197   & 286  \\
  Infected (\%)  & .1  & 85.4   & 88.0  & 69.2 & 89.7   & 93.5 \\
  \hline
\end{tabular}
\end{table*}

\section{Results}

\begin{figure}[tpb]
     \includegraphics[width=1.0\linewidth]{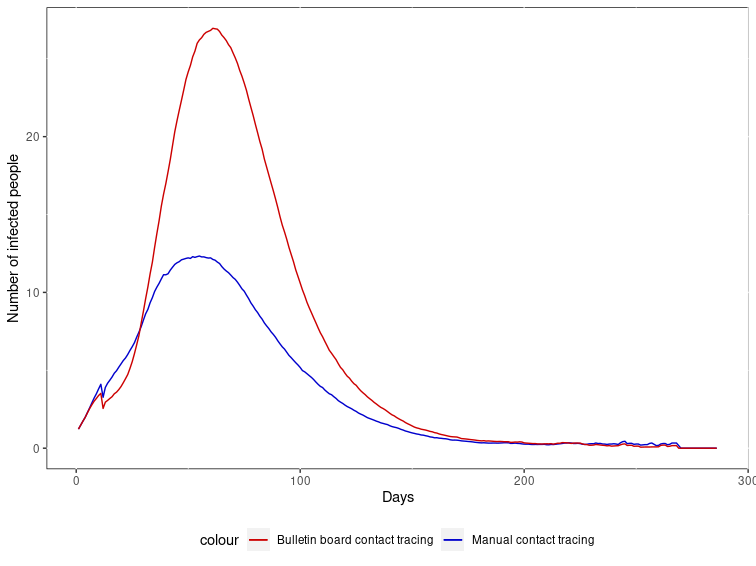}
     \caption{\textbf{Prevalence} of infection with manual and bulletin board contact tracing. N=1000. The graph shows mean values for each time period across all runs. The values are small because of the large number of runs that do not result in outbreaks but that are included in the divisor.}
     \label{prevalence}
\end{figure}

\begin{figure}[tpb]
     \includegraphics[width=1.0\linewidth]{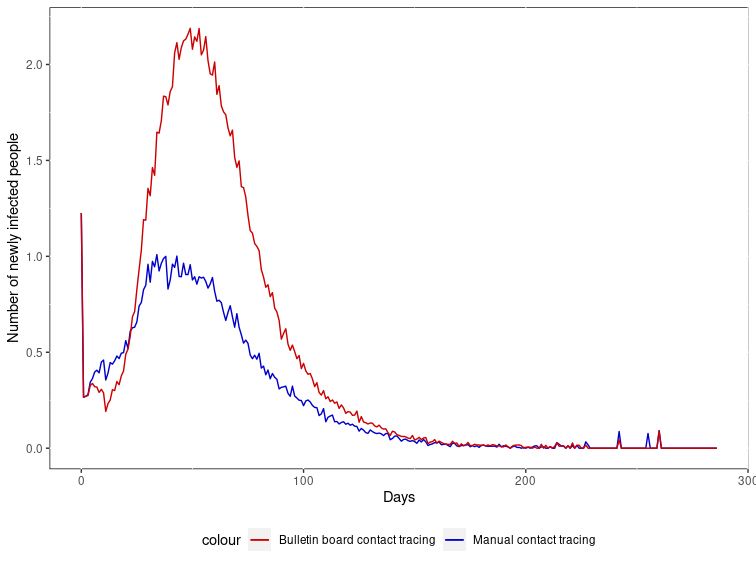}
     \caption{\textbf{Incidence} of infection with manual and bulletin board contact tracing. N=1000. The graph shows mean values for each time period across all runs. The values are small because of the large number of runs that do not result in outbreaks but that are included in the divisor.}
     \label{incidence}
\end{figure}

This paper was motivated by the question of how bulletin board contact tracing compares to manual contact tracing. To test this, we developed a discrete-time SEIR model of disease spread calbrated to what is known about SARS-CoV-2 (Table~\ref{m0calibrate}). We then combined this with models of symptomatic testing and the two types of contact tracing. 
In discrete-time simulations, compared to manual contact tracing, we find that bulletin board tracing does not significantly ($\alpha=0.05$) change the duration of the disease (Table~\ref{duration}). It also does not significantly reduce the percentage of the population that is infected (Table~\ref{infected}). Instead, it is effective at reducing the overall number of outbreaks. It also does not alter the overall $R_0$ (Table~\ref{results:rzero}) compared to manual contact tracing. Bulletin board tracing reduces prevalence initially (Figure~\ref{prevalence}), but then permits a higher prevalence than manual contact tracing. Incidence follows a similar pattern (Figure~\ref{incidence}).
All graphs in this section show the results of 1000 iterations of the model for each set of parameters. Each iteration consists of at most 1000 time periods. Parameter values are in Table~\ref{tab:parameters}.  

\begin{figure}[htpb]
 \begin{subfigure}{0.5\textwidth}
     \includegraphics[width=0.9\linewidth,height=5cm]{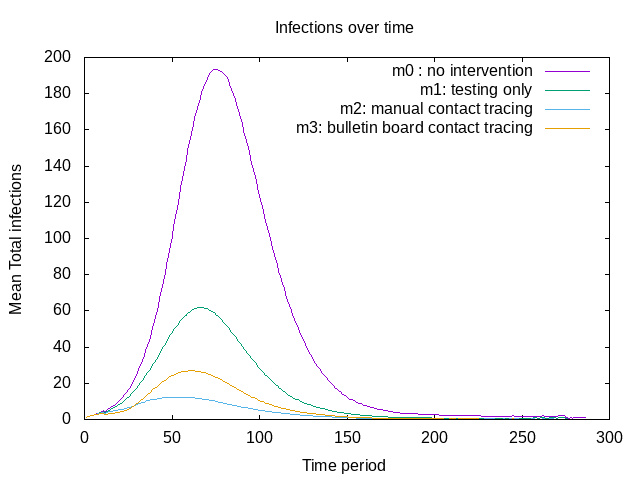}
     \caption{mean curves}
     \label{fig1means}
 \end{subfigure}
 \begin{subfigure}{0.5\textwidth}
     \includegraphics[width=0.9\linewidth,height=5cm]{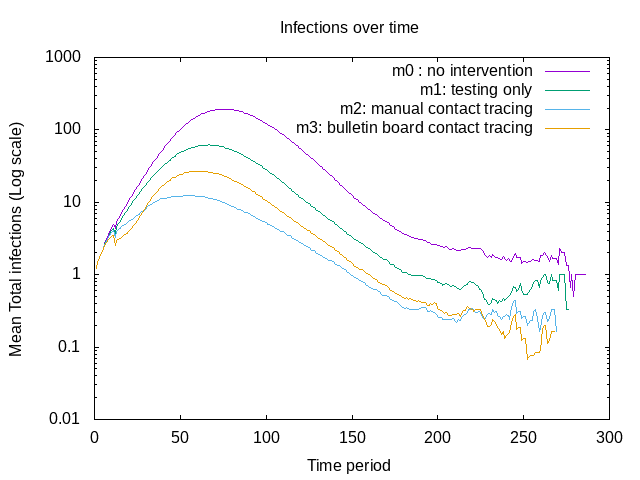}
     \caption{mean curves, Log scale}
     \label{fig2means}
 \end{subfigure}
     \caption{\textbf{Mean total infections over time for all models:} Mean total infection curves for 1000 runs of the model, with model density of 100 ( $N=1\times10^3$ and $L=1\times10^1$) and various parameters for other models.}
     \label{fig:means}
 \end{figure}

\begin{figure}[htpb]
     \includegraphics[width=0.9\linewidth]{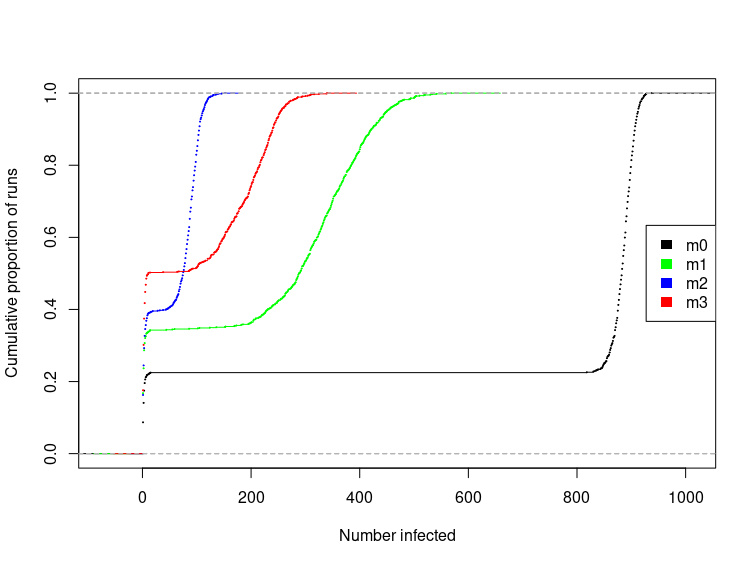}
     \caption{\textbf{The cumulative fraction} of runs by total number infected for all four models.}
     \label{fig:m3runs}
\end{figure}

\begin{table*}
\centering
\caption{\textbf{Results[Duration (days)]}: Results of comparing manual (m2) versus bulletin board (m3) contact tracing. 1000 runs with parameter values given in Table~\ref{tab:parameters}.}
\label{duration}
\begin{tabular}{@{}rrrrrrr@{}}
\hline
  & Min.  & 1st Qu. & Median & Mean [95\%CI]  & 3rd Qu. & Max.   \\
  \hline
  m2 & 11.00  & 13.00 & 119.00 &  95.26 [90.98-99.53] &  152.00 & 269.00 \\
  m3 & 11.00 &  11.00 &  39.00 &  87.34 [82.67-92.0]&  158.00 &  251.00 \\
\hline
\end{tabular}
\end{table*}

\begin{table*}
\centering
\caption{\textbf{Results[Infected (\%)]}: Results of comparing manual (m2) versus bulletin board (m3) contact tracing. 1000 runs with parameter values given in Table~\ref{tab:parameters}.}
\label{infected}
\begin{tabular}{@{}rrrrrrr@{}}
\hline
  & Min.  & 1st Qu. & Median & Mean [95\% CI]  & 3rd Qu. & Max.   \\
  \hline
 m2 &    0.1   &  0.3 &  7.5 &  5.6 [5.3-5.8] &  9.4 & 15.2  \\
   m3 & 0.1  &   0.2  &  1.1  &  9.9 [9.3-10.6] &  20.2  & 34.0 \\
\hline
\end{tabular}
\end{table*}

\begin{table*}
\centering
\caption{\textbf{Results[$R_0$]}: Results of comparing manual (m2) versus bulletin board (m3) contact tracing. 1000 runs with parameter values given in Table~\ref{tab:parameters}.}
\label{results:rzero}
\begin{tabular}{@{}lrrrrrr@{}}
\hline
  & Min.  & 1st Qu. & Median & Mean [95\% CI]  & 3rd Qu. & Max.   \\
  \hline
m0: Baseline       & 1.000 &  1.000 &  1.001 &  1.072 [1.059-1.086] &  1.117 &  1.492 \\
m1: Testing Only       &   1.000 &  1.000 &  1.002 &  1.054 [1.042-1.065] &  1.063 &  1.478 \\
m2: Manual Contact Tracing      & 1.000  & 1.000 &  1.002 &  1.040 [1.030-1.050] &  1.033 &  1.478 \\
m3: Buleltin Board Tracing      &   1.000 &  1.000 &  1.002 &  1.040 [1.031-1.050] &  1.037 &  1.471 \\
   \hline
\end{tabular}
\end{table*}

\subsubsection{m0: No Intervention}
This is the baseline scenario in each of the comparisons. We assume that there are ten locations, one thousand people in the population ($\rho=100$) and no heterogeneity.

\subsubsection{m1: Testing only}
This is the model of testing and subsequent isolation. We assume that 70\% of individuals isolate when asked to, or after they get a positive result.  We assume that 2 in 10 tests gives a false negative. Predictably, less of the population on average is infected in the model compared to m0.

\subsubsection{m2: Manual Contact Tracing}
 In the manual contact tracing scenario, we keep the testing and false negative testing rates the same as model m1. We assume that individuals can recall 50\% of all the people they contacted in the past 11 days, including strangers. And of these people, we assume that 50\% can be contacted after identification. We assume that the contact tracing instantiation is highly efficient and that it only takes one day between the positive test result and the contacts receiving a message from the contact tracer. Finally, we assume that 50\% of individuals cooperate with contact tracers. The results thow a significant improvement over testing alone both at reducing duration incidence and the effective reproduction number.

\subsubsection{m3: Bulletin Board Contact Tracing}
In the bulletin board contact tracing scenario, we keep the testing parameters the same as with manual contact tracing. We assume that individuals can recall 70\% of all the locations they visited in the past 11 days. This is higher than the rate of recalling individual contacts because it is easier to remember locations than people. Of the people who visited these locations at the same time, we assume that 50\% check the bulletin board at least once a day or have signed up to receive notifications. We assume that the bulletin board contact tracing implementation is highly efficient and that the message reaches the individuals the same day that it is posted. Finally, we assume that half of all individuals cooperate fully with the bulletin board system.
The bulletin board approach leads to a shorter mean duration (Table~\ref{duration}), although not significantly so as well as a significantly larger number of total infected on average (Table~\ref{infected}). The effective reproduction number remains the same, however. 

\section{Discussion}

Bulletin board contact tracing can improve manual contact tracing considering how easy it is to implement, the relative fewer resources it demands, and its nuanced privacy options.
However, as our models show, under certain conditions, bulletin board contact tracing can be comparable or even slightly worse than manual contact tracing.
Figure~\ref{fig:means} shows that around the 45-day mark, the curves for manual contact tracing and bulletin board tracing cross with the former trending downwards faster. We suspect that this reflects the difference in expected numbers of infected people due to a missed location versus a missed individual. This is likely because locations typically have more than one person in it. So, missing a location that one had visited that was infected rules out a larger number of people from receiving the message to isolate. In comparison, missing one person that was contacted rules out only one person from receiving the message to isolate. Thus, it is a limitation of the model that it does not adequately scale the probabilities of recall. It also represents a significant handicap in the head to head comparison of the two models of contact tracing.

We suspect this crossing is likely only when (i) the recall of locations is comparable to recall of contacts and (ii) the probability of a message reaching the recipient after they had been identified is about as good as, or slightly better than, a coin toss in both cases. This is not a realistic scenario in light of what we know from individual experiences. First, there are strong reasons to believe that our ability to remember locations is better than the ability to remember individuals (some of whom will be strangers). Second, if we visit locations with lots of people (such as concerts, or other super-spreading venues), we are even more likely to remember these locations. Third, technology to assist in identifying locations is far more accurate and ubiquitous than the technology to identify people. And finally, based on authors' experiences, people are more willing to share anonymous location-identifying information than individually identifying information.Taken together, a contact tracing approach that also takes into account location information, and uses a bulletin board approach is likely to be more effective than manual contact tracing alone in all but the most uncommon circumstances.

We have been as generous as we could to manual contact tracing given recent reports of serious problems with the approach, and the limitations of technological fixes to the approach. On the other hand, we have been relatively severe to the bulletin board approach by assuming that the key parameters are similar to manual contact tracing when everyday experience suggests otherwise. Our results, therefore, should be understood as sketching a lower bound on the gains that can be achieved through bulletin board contact tracing. There are multiple avenues for improving its performance significantly compared to what we have assumed, without incurring similar costs as manual contact tracing. It is worth highlighting that bulletin board approach can propogate messages much faster than the manual approach, and people are more likely to remember locations than people, particularly strangers. Compliance is also likely to be higher when there there is greater privacy by default.

There are several extensions of our model that immediately suggest themselves. One is the addition of a probability of staying home. Another is the addition of a false positive rate, in order to better model different testing regimes. However, it should be noted that the false positive rate matters to the model only when one uses asymptomatic testing. One could also generate the individual basic reproductive number from a probability distribution to introduce heterogeneity into the model. We have also used a lognormal distribution to sample the infectious period from, however this could be generalized to choose any other distribution.

\section{Conclusion}
Based on these results, classical contact tracing should, in the least, incorporate location information alongside individual information. If greater privacy options or efficiency is desired, a greater reliance on location-based tracing should be considered compared to manual contact tracing as classically implemented in most circumstances.
Bulletin board contact tracing provides an improved solution for the two main problems of classical contact tracing: reconstruction of the contact set for the infected individual and communicating with those who have been exposed. 
It improves contact set reconstruction by relying on individuals' memory of locations not of people; and by combining the location memories of the index individual as well as the exposed individuals. It improves on the message passing by similarly requiring only that the index person pass a message to a public bulletin board. 
The recipients also have only to know how to get messages from this board. The contact tracer, thus, does not need to know how to individually reach every contact.
However, the bulletin board approach makes these improvements at the cost of requiring greater cooperation with the contact tracing regime than classical contact tracing. This, as we have noted, may be easier to secure than classical contact tracing because individuals involved need to reveal a relatively negligible amount of personal information: they need only agree to the posting of an anonymous message associating location and time.
Based on our models, we suggest that policy makers should consider bulletin board based approaches to contact tracing in addition to manual approaches. 

\begin{acknowledgements}
  The authors are deeply indebted to the advice, guidance and generosity of Edward Kaplan.
  Sarah Lewis contributed the survey of recent contact tracing technologies.
  
\end{acknowledgements}

%
%

\bibliographystyle{style/spbasic}      


\bibliography{main}

\end{document}